\documentclass[runningheads]{llncs}
\usepackage{graphicx}
\begin{document}
\title{Approximate Solutions of a Kinetic Theory for Graphene}
%
\author{D.~B.~Blaschke\inst{1,2,3}
\and
V.~V.~Dmitriev\inst{4} \and
N.~T.~Gevorgyan\inst{5,6}
\and
B.~Mahato \inst{1}\and
A.~D.~Panferov\inst{4}
\and
S.~A.~Smolyansky\inst{4,7}
 \and
V.~A.~Tseryupa\inst{4}}
\authorrunning{D.~B.~Blaschke et al.}
%
\institute{Institute for Theoretical Physics, University of Wroclaw, 50-204 Wroclaw, Poland \and
Bogoliubov Laboratory for Theoretical Physics, Joint Institute for Nuclear Research, RU - 141980 Dubna, Russia \and
National Research Nuclear University (MEPhI), RU - 115409 Moscow, Russia \and
Saratov State University, RU - 410026 Saratov, Russia \and
A.I.~Alikhanyan National Science Laboratory (YerPhI), 0036 Yerevan, Armenia \and
Byurakan Astronomical Observatory, National Academy of Sciences, 0213 Byurakan, Armenia \and
Tomsk State University, RU - 634050 Tomsk, Russia \\
\email{david.blaschke@uwr.edu.pl; smol@sgu.ru}}

\maketitle              
\begin{abstract}
The effective mass approximation is analysed in a nonperturbative kinetic theory approach to strong field excitations in graphene ~\cite{Smolyansky:2019,Smolyansky:2020b}.
This problem is highly actual for the investigation of quantum radiation from graphene~\cite{Universe:2020}, where the collision integrals in the photon kinetic equation are rather complicated functionals of the distribution functions of the charge carriers.
These functions are needed in the explicit analytical definition as solutions of the kinetic equations for the electron-hole excitations.
In the present work it is shown that the suggested approach is rather effective in a certain range of parameters for the pulse of an external electromagnetic field. For example, the applicability condition of the approximation in the case of a harmonic field is
$\hbar \omega ^2 / (\sqrt 2 e E_0 v_F) < 1, $
were $v_F$ is the Fermi velocity.
In the standard massive quantum electrodynamics the usability of the analogical approximation is very narrow.

\end{abstract}
\section{Introduction}

It  is  well  known that the interaction  of  carriers in graphene with with an external electromagnetic field is nonanalytic in the coupling constant~\cite{Vozmediano:2010,Lewkowicz:2011}.  
This renders  applications of perturbation theory unjustified and stimulates the search for nonperturbative approaches. Paradigms are here exactly  solvable quantum field theory models~\cite{Gavrilov:2012,Klimchitskaya:2013}. 
However, these solutions are limited to narrow classes of external field models (constant electric field, Eckart's potential). Alternative cases are founded either on direct  application  of  the basic equations of motion~\cite{Ishikava:2010,Ishikava:2013} or on the nonperturbative  kinetic theory~\cite{Smolyansky:2019,Smolyansky:2020b,Universe:2020,Panferov:2017} constructed in analogy to the standard strong field QED~\cite{Grib:1994,Birula:1991,Schmidt:1998,Kluger:1998}.

A  higher level of description of the graphene excitations is connected with the investigation  of  different  mechanisms  of  radiation.  In the simplest   case   the   question  is  about  the emission  of  a quasiclassical electromagnetic  field  by the plasma currents in graphene~\cite{Bowlan:2014,Baudisch:2018}.
There is also the radiation  of a quantized field that is the result of the direct interaction of the charge carriers with  the photon field. 
A consistent realization of this approach is based on a truncation procedure of the
Bogoliubov-Born-Green-Kirkwood-Yvon  (BBGKY)  chain  of  equations  for  the correlation  functions of the $eh\gamma$ -- system~\cite{Universe:2020}. 
This leads to a closed  system  of  kinetic  equations  (KEs)  with collision integrals  (CIs)  of  non-Markovian type in the electron-hole and photon sectors and with the Maxwell equation for the acting inner (plasma)  field.  
As far as the evolution of the $eh$-plasma is accompanied by  high-frequency  quantum  (vacuum, as in the case of the electron - positron  plasma)  oscillations,  any solution of this KE system turns out  very susceptible to the selection not only of the model and the parameters of the external  field  but  also to the necessary roughening in the process of calculating the physical quantities.

Below  we  will  consider  this  problem  on  the  examples of two approximative methods  for the solution  of  the  basic  KEs describing the production of $eh$-pairs under the action of an external field: the low density  approximation~\cite{Schmidt:1999}  and  the  method  of  the  asymptotic decompositions~\cite{Mamaev:1979} (Sect. 2). 
The ideas for these approaches are borrowed from the standard QED. 
In the  considered  case of the massless theory, an essential role plays the  effective  mass  approximation~\cite{Berestetskii:1982}.  
The  results  of analytical calculations of the basic functions of the kinetic theory are compared among themselves and with the exactly solvable model (Sect. 3). 
The considered examples show that the introduction of the effective electromagnetic mass allows to obtain rather simple expansions for the distribution functions of the $eh$-plasma in graphene for time dependent electric fields with different pulse shape.

\section{The basic KE and its approximate solutions}

The basic KE for description of excitation and evolution of the $eh$-plasma in graphene under the action of an external quasiclassical spatially uniform time-dependent electric field with the vector potential $A^{\mu}=(0,A^{(1)}(t),A^{(2)}(t),0)$ in the Hamiltonian gauge $A^{(0)}=0$ can be written in the integro-differential form~\cite{Smolyansky:2019,Smolyansky:2020b,Universe:2020,Panferov:2017}
\begin{equation}
\label{eq1}
\dot{f}(\mathbf{p},t)=\frac{1}{2}\lambda (\mathbf{p},t)\int_{t_{0}}^{t}\lambda (\mathbf{p},t^{\prime })[1-2f(\mathbf{p},t^{\prime })]\cos \theta (\mathbf{p};t,t^{\prime })dt^{\prime }
\end{equation}
or as the equivalent system of ordinary differential equations
\begin{eqnarray}\label{eq2}
&&\dot{f}(\mathbf{p},t)=\frac{1}{2}\lambda (\mathbf{p},t)u(\mathbf{p},t),
\\ \nonumber
&&\dot{u}(\mathbf{p},t)=\lambda (\mathbf{p},t)\left[ 1-2f(\mathbf{p},t)%
\right] -\frac{2\varepsilon (\mathbf{p},t)}{\hbar }\nu (\mathbf{p},t),\ \dot{%
\nu}(\mathbf{p},t)=\frac{2\varepsilon (\mathbf{p},t)}{\hbar }u(\mathbf{p},t).
\end{eqnarray}
Here $f(\mathbf{p},t)$ is the distribution function of charge carriers introduced by taking into account the electroneutrality $f(\mathbf{p},t)=f_e(\mathbf{p},t)=f_h(-\mathbf{p},t)$. 
The excitation function $\lambda (\mathbf{p},t)$ in the low energy model is determined as
\begin{equation}\label{eq3}
\lambda (\mathbf{p},t)=\frac{ev_{F}^{2}[E^{(1)}(t)P^{2}-E^{(2)}(t)P^{1}]}{\varepsilon ^{2}(\mathbf{p},t)},
\end{equation}
where $v_{F}=10^6\ $ m/s is the Fermi velocity, $E^{(k)}=-\frac{1}{c}\dot{A}^{(k)}(t)$ is
the field strength $(k=1,2)$, $P^k=p^k+\frac{e}{c}A^k(t)$ is the quasi-momentum and $\varepsilon(\mathbf{p},t)=v_F \sqrt{P^2}$ is the quasi-energy. 
The electron charge is $-e$. 
Finally, the quantity $\theta (\mathbf{p};t,t^{\prime })$ in the KE (\ref{eq1}) is the phase,
\begin{equation}\label{eq4}
\theta (\mathbf{p};t,t^{\prime })=\frac{2}{\hbar }\int_{t^{\prime
}}^{t}dt^{\prime \prime }\varepsilon (\mathbf{p},t^{\prime \prime }).
\end{equation}

The KE (\ref{eq1}) is an integro-differential equation of the non-Markovian type with a fastly oscillating kernel. 
There is an integral of motion
\begin{equation}\label{eq5}
(1-2f)^2+u^2+v^2={\rm const},
\end{equation}
where the constant is fixed with the corresponding initial condition.

The KEs (\ref{eq1}), (\ref{eq2}) for the case of graphene were obtained in the works~\cite{Smolyansky:2019,Smolyansky:2020b,Universe:2020,Panferov:2017} by the Bogoliubov method of canonical transformations that can by realized in the massless $D=2+1$ theory in an explicit form. 
On the other hand the system (\ref{eq2}) can be reduced from the general system of twelve KEs in the standard QED~\cite{Aleksandrov:2020}.

The  massless low energy model of graphene with the  lightlike dispersion  law  $\varepsilon(\mathbf{p},t)$ leads 
to the absence of the critical field that is characteristic for massive QED and results in a specific feature of the momentum dependence of the excitation function $\lambda (\mathbf{p},t)$ (\ref{eq3}): 
this  function decreases in the ultraviolet area, $\lambda (\mathbf{p},t)\sim 1/P \rightarrow 0$  at  
$ P\rightarrow\infty$   and  is  singular  in  the infra-red area $\lambda  (\mathbf{p},t) \rightarrow\infty$ at $P\rightarrow0$. 
The last distinction also leads to a nonanalytic structure of the theory in its dependence on the coupling constant and to the absence of the standard perturbation theory.

Let  us  write  also the differential equation of the third order that is equivalent to the system of equations (\ref{eq2}),
\begin{equation}
\label{eq6}
\stackrel{...}{g}+\ddot{g}\left[\varepsilon \lambda \left(\frac{1}{\varepsilon \lambda}\right)^{\prime}-\frac{\dot{\lambda}}{\lambda} \right]+\dot{g}\left[ 4\varepsilon^2+
\lambda^2-\varepsilon \lambda\left(\frac{\dot{\lambda}}{\varepsilon \lambda^2}\right)^{\prime}\right]+
g \varepsilon \lambda \left(\frac{\lambda}{\varepsilon} \right)^{\prime}=0,
\end{equation}
where $g=1-2f$. In Eq. (\ref{eq6}) the $^{\prime}$ denotes  also the time derivative.

At the present time, an exact solution of the KEs (\ref{eq1}), (\ref{eq2}) is not known. 
However, below we will assume that the well-known exact solutions of the Dirac equation for a constant electric field and the Eckart potential are at the same time solutions of the KEs (\ref{eq1}), (\ref{eq2}). 
This assumption gives a basis for comparing these exact solutions with the known approximate solutions of the KEs (\ref{eq1}), (\ref{eq2}) and to construct then some new classes of approximate solutions.

In  order to estimate the effectivity of the approximate solutions, we  will compare them to the  exact  solutions  (analytical  and numerical). 
Such a comparison will be made on the level of an  integral macroscopic quantity. 
The number density of pairs $n(t)$ will be considered as the simplest quantity of such type,
\begin{equation}
\label{eq7}
n(t)=\frac{N_f}{(2\pi\hbar)^2}\int f(\mathbf{p},t)d\mathbf{p},
\end{equation}
where  $N_f=4$  is  the number of flavours. 
The integration allows here to smoothen out some insignificant details in the momentum dependence of
the distribution functions in different approximations.

As the next step, we will consider two approximate methods of solving the KEs (\ref{eq1}), (\ref{eq2}).
To this end, we will consider the case of a linearly polarized electric field
$A^{(1)}=0,\ A^{(2)}(t)=A(t)$.

\subsection{Low density approximation} 
This approximation corresponds to the limit $f\ll 1$ in the r.h.s. KE (\ref{eq1}). It was introduced  in the strong field vacuum production of charged particles~\cite{Schmidt:1999} and was used many times in strong field QED and in the kinetic theory of excitations in graphene. 
It leads to the quadrature formula~\cite{Smolyansky:2019,Smolyansky:2020b}
\begin{equation}\label{eq8}
f(t)=\frac{1}{4}\left[ \int_{-\infty }^{t}dt^{\prime }\lambda _{s}(t^{\prime
})\right] ^{2}+\frac{1}{4}\left[ \int_{-\infty }^{t}dt^{\prime }\lambda
_{c}(t^{\prime })\right] ^{2},
\end{equation}
where
\begin{equation}
\label{eq9}
\lambda _{c}(t)=\lambda (t)\cos \theta (t,-\infty ),\ \ \lambda
_{s}(t)=\lambda (t)\sin \theta (t,-\infty ).
\end{equation}
In particular, it follows from Eq. (\ref{eq8}), that $f(\mathbf{p},t)\geq 0$.

In the low-density approximation, the two last KEs of the system (\ref{eq2}) separate from it,
\begin{equation}\label{eq10}
\dot{u}=\lambda -\frac{2\varepsilon}{\hbar} v,\ \dot{v}=\frac{2\varepsilon}{\hbar} u.
\end{equation}
Then the distribution function $f$ can be found from the first equation of the system (\ref{eq2}) alone.

The system of equations (\ref{eq10}) corresponds to the ordinary differential equation,
\begin{equation}\label{eq11}
\frac{\hbar}{2\varepsilon}\frac{d}{dt}\left( \frac{\hbar}{2\varepsilon} \frac{du}{dt} \right)
+u- \frac{\hbar}{2\varepsilon}\frac{d}{dt}\left( \frac{\hbar\lambda}{2\varepsilon} \right)=0
\end{equation}
or
\begin{equation}\label{eq12}
{\cal D}^2 u + u - {\cal D} \left( \frac{\hbar\lambda}{2\varepsilon} \right) =0,\ {\cal D}=\frac{\hbar}{2\varepsilon}\frac{d}{dt}.
\end{equation}

\subsection{Effective electromagnetic mass approximation} 
This  approximation  was  used  already  in the case of the harmonic model  of  an external field in the analysis 
of the radiation effects in the $e^-e^+$ plasma~\cite{Fedotov:2020} and in the $eh$ plasma in graphene~\cite{Universe:2020}. 
Below we will consider a generalization of this approach to other models of the external field.

The idea of the method is that the time dependent value of the square of the kinetic momentum $P^2(t)$ 
in the definition of the quasienergy $\varepsilon(\mathbf{p},t)$ gets substituted by corresponding time average 
$\prec P^2(t) \succ$, where the symbol $\prec ... \succ$ means the averaging procedure over a characteristic time of the external field. In the case of the linearly polarized electric field we obtain
\begin{equation}
\label{eq13}
\prec P^2_2(t) \succ = p^2_2 +(e/c)^2 \prec A^2(t) \succ ,
\end{equation}
if $\prec A(t) \succ =0$. Implying the substitution
\begin{equation}\label{eq14}
P^2_2(t) \rightarrow \prec P^2_2(t) \succ,
\end{equation}
in the square of quasienergy $\varepsilon^2(\mathbf{p},t)=v^2_F[p^2_1+P^2_2(t)]$, one can introduce the longitudinal (with respect to external field) effective electromagnetic mass
\begin{equation}\label{eq15}
(e/c)^2 \prec A^2(t) \succ =m^2_{\ast} v^2_F,
\end{equation}
or
\begin{equation}\label{eq16}
m_{\ast }^{2}=\frac{e^{2}}{c^{2}v_{F}^{2}}\frac{1}{2T}%
\int_{-T}^{T}dtA^{2}(t).
\end{equation}
Thus, this approximation corresponds to change
\begin{equation}\label{eq17}
\varepsilon(\mathbf{p},t)\rightarrow\varepsilon_{\ast}(\mathbf{p})=v_F\sqrt{p_1^2+(p_2^2+m^2_{\ast}v^2_F)}.
\end{equation}

The appearance of the longitudinal mass is a reflection of the anisotropy of the system stipulated by the presence of the external field and leads to a reduction of the mobility of charge carriers along the direction of the action of the external field. In the limiting case $p_2 \ll m_{\ast}v_F$, we obtain a strong anisotropic momentum dependence of the quasienergy,
\begin{equation}\label{eq18}
\varepsilon_{\ast}(\mathbf{p})=v_F\sqrt{m^2_{\ast}v^2_F + p_1^2}.
\end{equation}

The approximation of the effective mass (\ref{eq14}), (\ref{eq15}) is valid in field models with
square integrable functions $A(t)$ only.

The transition amplitude (\ref{eq3}) in the effective mass approximation in the linearly polarized external field will be 
\begin{equation}
\label{eq19}
\lambda_{\ast}(\mathbf{p},t)=-\frac{ev^2_F p_1}{\varepsilon^2_{\ast}(\mathbf{p})}E(t)\equiv \Lambda (\mathbf{p}) E(t),
\end{equation}
where $p_1=\cos \varphi$ and $\varphi$ is the polar angle.

The problem of evaluating the distribution function is now brought to the calculation of the integral
\begin{equation}
\label{eq20}
J(\mathbf{p},t)=\int_{-\infty}^{t}dt^{\prime} E(t^{\prime}) \cos \theta_{\ast}(\mathbf{p}; t^{\prime}, -\infty),
\end{equation}
where $\theta_{\ast}(\mathbf{p}; t^{\prime}, -\infty)$ is defined by the relation (\ref{eq4}) with the replacement 
$\varepsilon(\mathbf{p},t)\rightarrow \varepsilon_{\ast}(\mathbf{p})$. 
The analogous integral with the replacement $\cos \theta_{\ast}\rightarrow \sin \theta_{\ast}$ is equal to zero, if $E(t)=E(-t)$. The distribution function (\ref{eq8}) will then be 
\begin{equation}\label{eq21}
f(\mathbf{p},t)=\frac{1}{4}\Lambda^2(\mathbf{p})J^2(\mathbf{p},t).
\end{equation}

According   to   the   relations  (\ref{eq19})  and  (\ref{eq21}), the  anisotropy  of  the distribution function 
$\sim \cos^2 \varphi$ is universal and does not depend on the selection of the external field model. 
Another distinctive feature of the amplitude (\ref{eq19}) is the singular slit of the surfaces $\lambda(\mathbf{p},t)$ 
on the plane $(p_1,p_2)$ along the axis $p_1=0$ (or $\varphi=\pi/2$) for $p_2\neq 0$, i.e. 
$\lambda(\mathbf{p},t)|_{p_1=0}=0,$ and $ \lambda(\mathbf{p},t)|_{p_1\neq0}\neq0$. 
As it follows from the KEs (\ref{eq1}) and (\ref{eq2}), this peculiarity is reproduced also in the distribution function, $f(\mathbf{p},t)|_{p_1=0}=0$. 
This means that quasiparticles are not created in the directions of the external field action in the strict sense. 
Fig. 1 demonstrates this on the example of the Sauter impulse. 
This slit is evident in the figures shown below in the case of the massless version of the theory. 
The introduction of a mass results in a widening of this slit and in the appearance of the energy gap. 
Let us remark, that the exact solution of the problem \cite{Klimchitskaya:2013} in the case of the Sauter pulse 
field has the singular line $p_1=0$ in  the case of the linearly polarized external field. 
The presence of this infinitely thin slit  is  not  reflected  in calculations of the integral "observable"
macroscopical averages of the type of the pair number density (\ref{eq7}).

\begin{figure}[!h]
\centering
\includegraphics[width=12 cm]{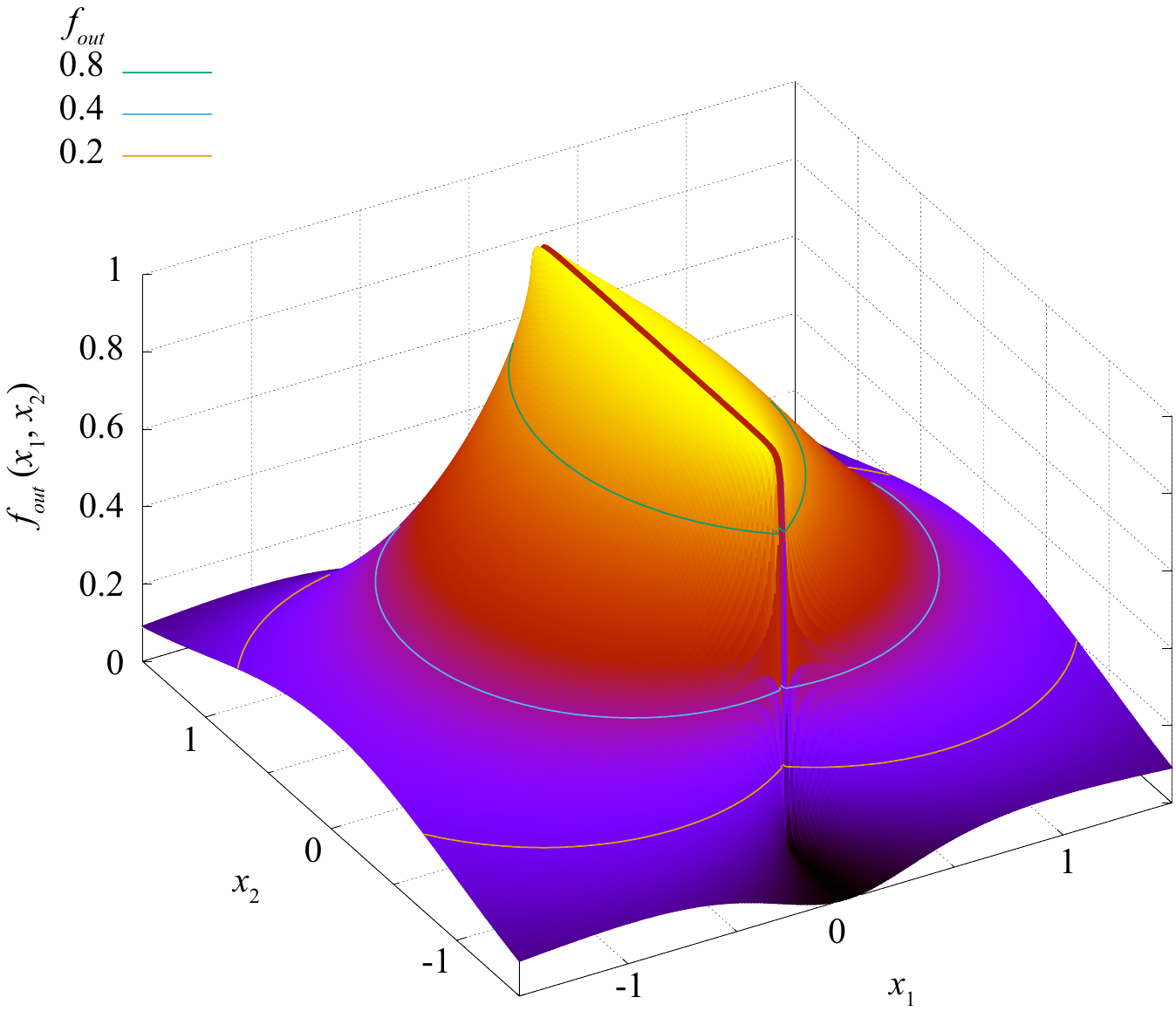}
\caption{The residual distribution function $f_{out}(x_1,x_2)$ for the Sauter field model with $E=1000\ {\rm V/cm}, \kappa = 10^{12}\ {\rm Hz}$, where $x_{1,2}=p_{1,2}/m_\ast v_F$ are dimensionless momenta.}
\label{F1}
\end{figure}

\subsection{Method of asymptotic decompositions} 

This method is adopted from the standard strong field QED~\cite{Mamaev:1979}.
We consider now the dimensionless excitation amplitude in the exact case (\ref{eq3})
\begin{equation}\label{eq22}
\Lambda(\mathbf{p},t)=e\hbar v^2_F E(t)p^1/2\varepsilon^3(\mathbf{p},t)
\end{equation}
and in the effective mass approximation
\begin{equation}\label{eq23}
\Lambda_{\ast}(\mathbf{p},t)=e\hbar v^2_F E(t)p^1/2\varepsilon^3_{\ast}(\mathbf{p},t).
\end{equation}
In contrast to $\Lambda(\mathbf{p},t)$, the amplitude $\Lambda_{\ast}(\mathbf{p},t)$ (\ref{eq23}) is limited everywhere,
\begin{equation}\label{eq24}
\Lambda_{\ast}(\mathbf{p},t)\leq \Lambda_{\ast}^{max}=\frac{e\hbar E_0}{3\sqrt{3}m^2_{\ast}v^3_F},
\end{equation}
where $\Lambda_{\ast}^{max}$ is the maximal value of the amplitude (\ref{eq23}) in the point of time where $E(t)=E_0$ (see Fig.1).

In order to clarify the physical meaning of the parameter $\Lambda_{\ast}^{max}$ given in (\ref{eq24}), let us consider the case of a harmonic field, where the momentum of the electromagnetic field is equal to 
$p_A=(e/c)A(t)\sim eE_0/\omega$. 
It corresponds to the contribution of the electromagnetic field in the quasienergy 
$\varepsilon_A=v_F p_A=e E_0 v_F/\omega$. 
Then the relation (\ref{eq24}) can be rewritten as
\begin{equation}
\label{eq25}
\Lambda_{\ast}^{max}=\frac{2\hbar \omega^2}{3\sqrt{3} e E_0 v_F}\sim \frac{\hbar \omega}{\varepsilon_A},
\end{equation}
which corresponds to the ratio of the energy of the absorbed quant of external field to the
part of energy of quasiparticle acquired as a result of acceleration in this field.

In the case of a sufficiently large low-frequency external field, $\hbar \omega \ll eE_0 v_F$, one can search a solution of the KE system (\ref{eq2}) by means of an asymptotic decomposition of the functions $f,\ u,\ v$ for the small parameter $\Lambda_{\ast}(\mathbf{p},t)\ll\Lambda_{\ast}^{max}\ll 1$,
\begin{equation}
\label{eq26}
f=\sum_{n=0}f_n, \quad u=\sum_{n=0}u_n,\quad v=\sum_{n=0}v_n.
\end{equation}
Substituting these decompositions in the KE system (\ref{eq2}) and equating the contributions of the same orders, 
one can obtain the leading terms of the asymptotic decompositions (\ref{eq26}) as
\begin{eqnarray}\label{eq27}
f_4 &=& \frac{1}{4}\Lambda^2_{\ast}(\mathbf{p},t)=\frac{e^2 v^4_F E^2(t) p_1^2}{16\varepsilon^6_{\ast}},\\ \nonumber
u_3 &=& \frac{1}{2\varepsilon_{\ast}}\dot{\Lambda}_{\ast}(\mathbf{p},t)=\frac{e v^2_F \dot{E}(t)p_1}{4\varepsilon^4_{\ast}},\quad v_2=\Lambda_{\ast}(\mathbf{p},t).
\end{eqnarray}

Formally, these expressions have the same form as the analogous results in standard QED which were obtained in the framework of the asymptotic decompositions of the functional series in $E_0/E_c\ll1$, where $E_c=m^2c^3/e\hbar$ is the critical field. In a similar way one can obtain the post-leading terms.

The obtained asymptotic solutions of the KE (\ref{eq2}) can be used for estimating the convergences of the integral macroscopical physical values (e.g., the densities of the conduction and polarization currents an so on) and also in analytical calculations in theory of radiation and other transport phenomena.

Let us consider now the realization of the effective mass approximation for different external field models.

\section{Approximate solutions of KEs for different external field models in graphene}

\subsection{The Sauter pulse} 

The field of this pulse is given by
\begin{equation}\label{eq28}
A(t)=-(cE_0/\kappa )\tanh \kappa t,\ E(t)=E_0/ \cosh ^2 \kappa t~.
\end{equation}
It is a classical example of the external field model leading to an exact solution of the basic equations of motion
of QED~\cite{Grib:1994}. 
The analogous solution for the massless graphene model was obtained in the work~\cite{Klimchitskaya:2013}.

The effective electromagnetic mass (\ref{eq16}) in this model is
\begin{equation}\label{eq29}
m_{\ast}=eE_0/v_F\kappa.
\end{equation}
The corresponding integral (\ref{eq20}) is
\begin{equation}\label{eq30}
J(\mathbf{p},t)=\frac{E_0}{\kappa}\int_{-\infty}^{\kappa t} dx \frac{\cos[\Omega(\mathbf{p})/\kappa]x}{\cosh^2x},
\end{equation}
where $\Omega(\mathbf{p})=2\varepsilon_{\ast}(\mathbf{p})/\hbar$. The  corresponding  distribution function $f(\mathbf{p},t)$ is defined then according   to   the  Eq.  (\ref{eq21}).  The  vacuum  polarization  functions $u(\mathbf{p},t)$ and $v(\mathbf{p},t)$ can be restored then with help of the Eqs. (\ref{eq2}). In the asymptotic limit $t\rightarrow\infty$ it follows that
\begin{equation}\label{eq31}
J_{out}(\mathbf{p})=\frac{2E_0}{\kappa}\frac{\pi \Omega(\mathbf{p})/2\kappa}{\sinh [\pi \Omega (\mathbf{p})/2\kappa]}.
\end{equation}
Then the distribution function (\ref{eq21}) in the out-state will be
\begin{equation}\label{eq32}
f_{out}(\mathbf{p})=\left[ \frac{ e E_0 v^2_F p_1}{\kappa \varepsilon^2_{\ast}(\mathbf{p})}\frac{\pi \Omega(\mathbf{p})/2\kappa}{\sinh [\pi \Omega (\mathbf{p})/2\kappa]} \right]^2.
\end{equation}
The corresponding expression for the pair number density (\ref{eq7}) written in terms of the dimensionless 
momentum $x_k = p_k/m_{\ast}v_F$ is 
\begin{equation}\label{eq33}
n_{out}=\frac{N_f}{\pi}\left( \frac{\eta^2 \kappa}{8\pi v_F} \right)^2 \int_{0}^{\infty} \frac{dx\ x^3}{1+x^2}\frac{1}{\left[ \sinh \left( \frac{\eta}{2} \sqrt{1+x^2} \right) \right]^2},
\end{equation}
where
\begin{equation}\label{eq34}
\eta =\frac{2\pi m_{\ast} v^2_F}{\hbar \kappa}=\frac{2\pi e E_0 v_F}{\hbar \kappa^2}.
\end{equation}

The point $\eta=1$ separates two domains: 
the domain $\eta<1$, where the energy of quasiparticles acquired in the external field per unit of time $e E_0 v_F/\kappa$ is less than the energy of an absorbed quant of the field $\hbar\kappa$, and the domain $\eta >1$, where the field acceleration mechanism dominates.
\begin{figure}
\includegraphics[width=\textwidth]{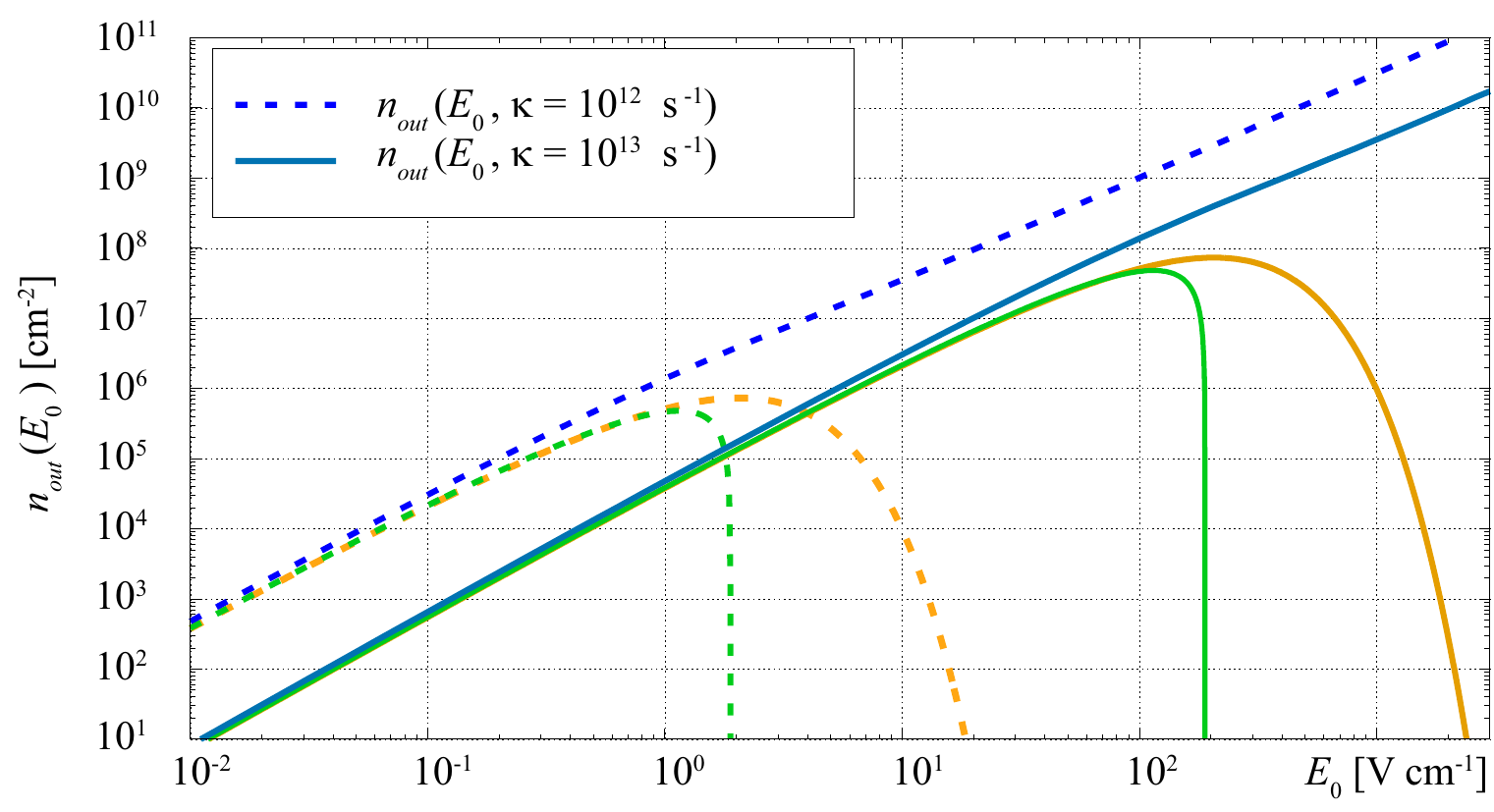}
\caption{Sauter field model for two values $\kappa = 10^{12}\ {\rm s}^{-1}$ and  $\kappa = 10^{13}\ {\rm s}^{-1}$. Blue lines are numerical solutions of KE (\ref{eq1}), orange lines are approximate solutions (\ref{eq33}), green lines are approximate solutions (\ref{eq35}).}
\label{F2}
\end{figure}

Some results of the numerical comparison of the exact and approximate (\ref{eq34}) dependencies $n_{out}(\eta)$  are given in Fig. 2. 
From here it follows that the effective electromagnetic mass approximation is valid in the domain $\eta \leq 1$ 
and can be dubbed "slow switching" with $\kappa \gg (2\pi e E_0 v_F/\hbar)^{1/2}$. 
Assuming that $\eta \ll 1$ on the r.h.s. of Eq. (\ref{eq33}), one can obtain
\begin{equation}\label{eq35}
n_{out}=-\frac{N_f}{\pi}\left( \frac{e E_0}{2\hbar \kappa}\right)^2 \left[ 1+ \frac{1}{2} \ln \frac{1}{24} + \ln \eta\right], \quad \eta \ll 1.
\end{equation}
This corresponds to the result obtained in \cite{Klimchitskaya:2013}.

\subsection{The Gaussian pulse} 
This field model
\begin{equation}\label{eq36}
A(t)=-\sqrt{\frac{\pi}{2}} c E_0\tau  {\rm erf} \left( \frac{t}{\sqrt{2}\tau} \right),\ E(t)= E_0 \exp (-t^2/2\tau ^2)
\end{equation}
results in the effective mass
\begin{eqnarray}\label{eq37}
m_{\ast}&=&\frac{e E_0  \tau \pi^{1/4}}{ v_F } \left( \sqrt{2}\ {\rm erf} (1/\sqrt{2})e^{-1/2}
+\sqrt{\pi}\ {\rm erf}^2 (1/\sqrt{2})/2 -{\rm erf}(1)\right)^{1/2}\nonumber\\
&\sim& 0.5257\ \frac{e E_0  \tau}{ v_F}.
\end{eqnarray}
The distribution function will be
\begin{equation}\label{eq38}
f(\mathbf{p},t)=\frac{1}{2}\left[\frac{e E_0 v^2_F \tau p_1}{\varepsilon^2_{\ast}(\mathbf{p})}\  I\left(\frac{t}{\sqrt{2}\tau},\sigma \right) \right]^2,
\end{equation}
where $\sigma=\Omega(\mathbf{p})\tau$ and
\begin{equation}\label{eq39}
I\left(\frac{t}{\sqrt{2}\tau},\sigma \right) =\int_{0}^{t/\sqrt{2}\tau}dx \cos (\sqrt{2}\sigma x) e^{-x^2}.
\end{equation}
A simple result follows from Eq. (\ref{eq38}) and (\ref{eq39}) in the asymptotic case $t\rightarrow\infty$,
\begin{equation}
\label{eq40}
f_{out}(\mathbf{p})=\frac{\pi}{2}\left\{ \frac{e E_0 \tau v_F^2 p_1}{\varepsilon^2_{\ast}(\mathbf{p})} \exp \left[-\frac{2\varepsilon^2_{\ast}\tau^2}{\hbar^2}\right] \right\}^2.
\end{equation}
From here one can find after simple calculations the pair number density in the out-state,
\begin{equation}\label{eq41}
n_{out}=-N_f\left( \frac{e\tau E_0}{4\hbar }\right)^2 \left[ e^{-\xi^2}+(1+\xi^2){\rm Ei} (-\xi^2) \right],
\end{equation}
where $\xi=2 m_{\ast} v^2_F\tau/\hbar$ and ${\rm Ei}(-\xi^2)$ is the exponential integral function.
This result in the region $\xi \ll 1$ corresponds to Eq. (\ref{eq35}) for the case of the Sauter pulse.
\begin{figure}
\includegraphics[width=\textwidth]{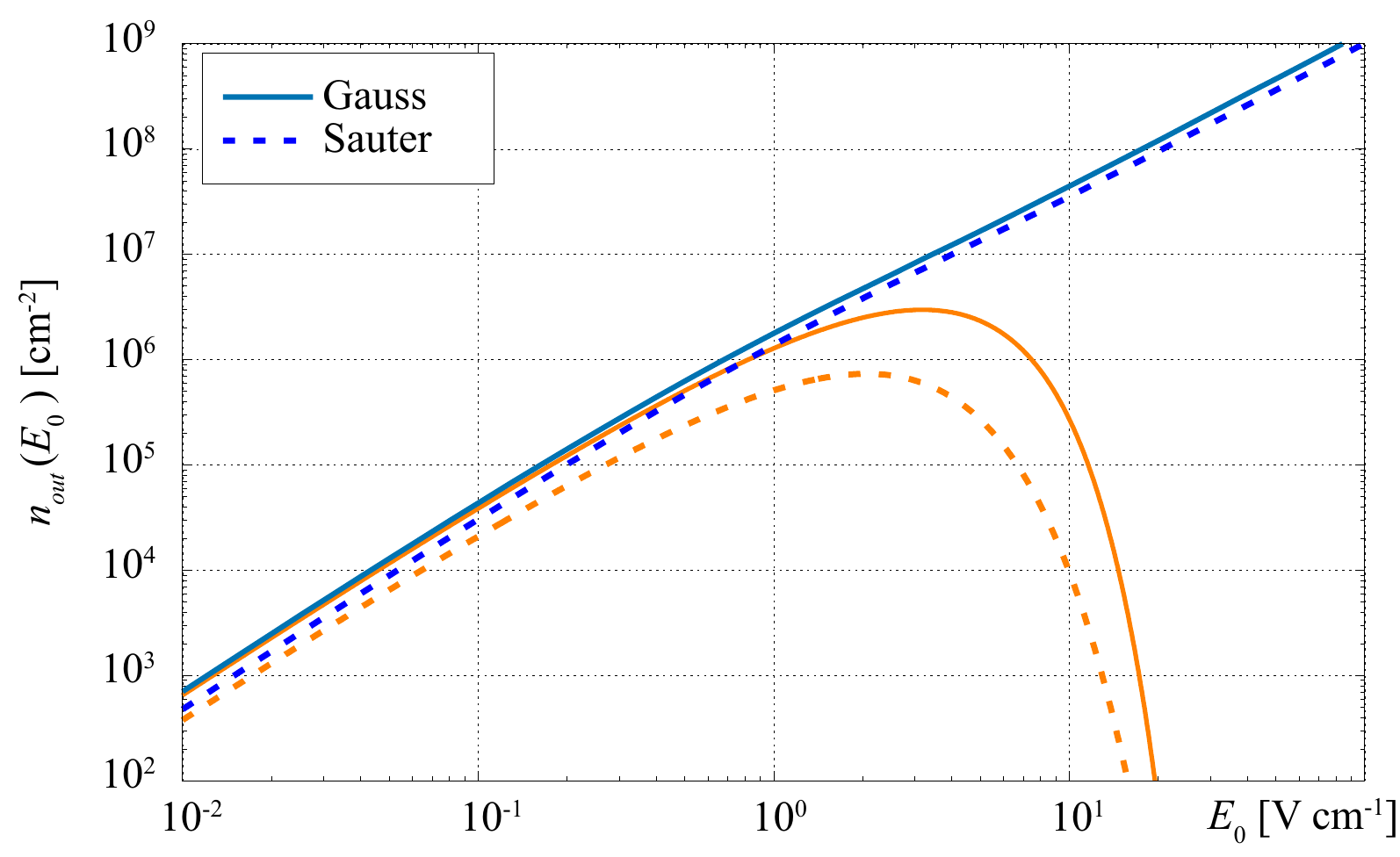}
\caption{The pair number densities $n_{out}$ for the numerical (blue lines) and approximate (orange lines) solutions for the Sauter ($\kappa=10^{12}\ \rm{s}^{-1}$) and Gauss  ($\tau=10^{-12}\ \rm{s}$) pulses.}
\label{F3}
\end{figure}
In Fig. 3 we compare the behaviour of the pair number densities $n_{out}$ for the exact and approximate solutions for the Sauter ($\kappa=10^{12}\ \rm{s}^{-1}$) and Gauss  ($\tau=10^{-12}\ \rm{s}$) pulses.

\subsection{The harmonic  field  model} 
The harmonic field model
\begin{equation}\label{eq42}
A(t)=-(cE_0/\omega) \sin \omega t,\ E(t)=E_0 \cos \omega t;
\end{equation}
corresponds to the effective mass
\begin{equation}\label{eq43}
m_{\ast}=eE_0/\sqrt{2}v_F\omega.
\end{equation}
The distribution function in this field model was obtained in the work~\cite{Universe:2020},
\begin{equation}\label{eq44}
f(\mathbf{p},t)=f^{(0)}(\mathbf{p})+f^{(2)}(\mathbf{p},t),
\end{equation}%
where the function
\begin{equation}\label{eq45}
f^{(0)}(\mathbf{p})=\frac{(e\hbar E_{0}v_{F}^{2}p_{1})^{2}(4\varepsilon _{\ast
}^{2}+\hbar ^{2}\omega ^{2})}{8\varepsilon_{\ast }^{4}(4\varepsilon _{\ast }^{2}-\hbar ^{2}\omega
^{2})^{2}}
\end{equation}%
corresponds to a stationary background distribution while the function
\begin{equation}\label{eq46}
f^{(2)}(\mathbf{p},t)=\frac{(e\hbar E_{0}v_{F}^{2}p_{1})^{2}}{8\varepsilon_{\ast }^{4}(4\varepsilon
_{\ast }^{2}-\hbar ^{2}\omega ^{2})}\cos 2\omega t
\end{equation}%
corresponds to the breathing mode on the doubled frequency of the external field.
The residual functions $u,\ v$ can be reconstructed using Eqs. (\ref{eq44})-(\ref{eq46}) and the KE system (\ref{eq2})
\begin{equation}
\label{eq47}
u(\mathbf{p},t)=\frac{e\hbar^2 v_{F}^{2}p_{1}}{\varepsilon_{\ast }^{2}(4\varepsilon
_{\ast }^{2}-\hbar ^{2}\omega ^{2})}\dot{E} (t),
\end{equation}
\begin{equation}\label{eq48}
v(\mathbf{p},t)=\frac{2e\hbar^2 v_{F}^{4}p_{1}}{\varepsilon_{\ast }(4\varepsilon
_{\ast }^{2}-\hbar ^{2}\omega ^{2})} E (t).
\end{equation}
The distribution function (\ref{eq44})-(\ref{eq45}) corresponds to the first and third
harmonics of the current density and radiation spectrum of the plasma
oscillations~\cite{Universe:2020}.
These results are valid in the case
\begin{equation}\label{eq49}
\hbar \omega ^2 / (\sqrt 2 e E_0 v_F) < 1.
\end{equation}
This  limitation holds also  for  other field models, if the quantity $\omega$ is interpreted as the corresponding characteristic frequency of the field alteration.

A general  feature  of  the two outlined approximate approaches is the  $E^2$ - proportionality of all the resulting distribution functions, $f(\mathbf{p},t)\sim(eE_0)^2$. 
This feature was obtained in the work~\cite{Kravtcov:2017} on  the  basis of an analysis of the numerical solutions 
of the corresponding KEs in standard QED, see also \cite{Blaschke:2019pnj}.

The effectiveness  of  the  low  density  approximation in the standard strong field  QED  for  rather  weak  fields  $E_0  \ll  E_c$  has been investigated  in  the  work~\cite{Fedotov:2011}. 
The additional introduction of the  effective mass approximation results herein a strong restriction of the domain of applicability of the method.

\section{Conclusion}
In  the  present  work  we have obtained some simple and rather universal results of approximate solutions for the  distribution functions of charge carriers in graphene based on nonperturbative KEs. 
This was achieved  by a combination of the low density approximation and the concept of an effective electromagnetic mass. 
Such an approach is effective for a rather wide  classes  of external  field models with the
parameters limited by the relation for the top (\ref{eq49}) of the characteristic frequency  $\omega$ of the external field.

The   considered   approximation   is  of  particular  interest  for the investigation of such complicated nonlinear single-photon effects in graphene as the emission (absorption) and annihilation (photoproduction)~\cite{Universe:2020}  and the more complex two-photon processes. 
Such kind of nonlinear phenomena in graphene became accessible  for experimental verification recently, see~\cite{Bowlan:2014,Baudisch:2018}.

\subsection*{Acknowledgements}
The work of D.B. was supported by the Russian Federal Program "Priority-2030". 
N.G. received support from Volkswagen Foundation (Hannover, Germany) under collaborative research grant 
No. 97029. B.M. acknowledges a stipend from the International Max Planck Research School for "Many-Particle Systems in Structured Environments" at the Max-Planck Institute for Physics of Complex Systems (Dresden, Germany).

\end{document}